\documentclass[thmsa,11pt]{article}
\begin{document}

\title{Irregular Hamiltonian Systems}

\author{Olivera Mi\v {s}kovi\'{c}$^{1,2}$ and Jorge Zanelli$^{1}$\\[2mm]
$^{1}$Centro de Estudios Cient\'{\i }ficos (CECS), Casilla
1469, Valdivia, Chile.\\ $^{2}$Departamento de F\'{\i }sica,
Universidad de Santiago de Chile,\\ Casilla 307,\ Santiago 2, Chile.
\\[2mm]
\texttt{olivera@cecs.cl, jz@cecs.cl}}
\date{}
\maketitle

\begin{abstract}
Hamiltonian systems with linearly dependent constraints (irregular
systems), are classified according to their behavior in the vicinity of the
constraint surface. For these systems, the standard Dirac procedure is not
directly applicable. However, Dirac's treatment can be slightly modified to
obtain, in some cases, a Hamiltonian description completely equivalent to
the Lagrangian one. A recipe to deal with the different cases is provided,
along with a few pedagogical examples.
\end{abstract}


\section{Introduction}

Dirac's Hamiltonian analysis provides a systematic method for finding the
gauge symmetries present in a theory. The analysis identifies and classifies
the constraints, which are local functions of the phase space coordinates.
Consistency requires that the constraints be preserved in during the
evolution (for the review of the Hamiltonian analysis see Ref.\cite{Dirac}--%
\cite{Chitaia-Gogilidze-Surovtsev}). However, if the constraints are not
functionally independent, then Dirac's procedure is not applicable. The test
of functional independence are the so-called regularity conditions, and
those systems which fail the test are said to be \emph{irregular}.

Irregular systems are not necessarily intractable or exotic. A simple
example is a relativistic massless particle ($p^{\mu }p_{\mu }=0$), which is
irregular at the origin of momentum space ($p^{\mu }=0$). This point in
phase space is exceptional, as it is unclear whether this would be an
observable state for a photon, say. On the other hand, we know the
configuration $p^{\mu }=0$ to be a very important one: the ground state.
There are other physical circumstances in which regularity is violated, and
not only for isolated states but on large portions of the region in phase
space where the system evolves. Chern-Simons theories in $2n+1$ spacetime
dimensions are examples where, for some initial configurations, regularity
can fail at all times and one is forced to live with this problem.

Here we discuss the possible ways in which the constraints can fail the test
of functional independence, and how the Hamiltonian treatment of Dirac must
be modified in each case.


\section{Regularity Conditions}

If we call $z^{i}\equiv (q,p)$ ($i=1,\ldots ,2n$) the coordinates in the
phase space $\Gamma $, the constraints $\phi ^{r}(z)\approx 0$ ($r=1,...R)$
define the \emph{constraint surface} $\Sigma $ given by
\begin{equation}
\Sigma =\left\{ \bar{z}\in \Gamma \mid \phi ^{r}(\bar{z})=0\;\left(
r=1,\ldots ,R\right) \left( R\leq 2n\right) \right\} .
\end{equation}

\textbf{Regularity Conditions}\emph{\ }(RCs)\emph{: The constraints }$\phi
^{r}\approx 0$\emph{\ are regular if and only if their small variations }$%
\delta \phi ^{r}\emph{,}$ \emph{evaluated} \emph{on }$\Sigma \emph{,}$\emph{%
\ are }$R$\emph{\ linearly independent functions of }$\delta z^{i}$\emph{.}%
\medskip

To first order in $\delta z^{i}$, the variation of the constraints are $%
\delta \phi ^{r}=J_{i}^{r}\delta z^{i}$, where $J_{i}^{r}\equiv \left. \frac{%
\partial \phi ^{r}}{\partial z^{i}}\right| _{\Sigma }$. Consequently, the
RCs can also be defined as \cite{Henneaux-Teitelboim}:\medskip

\emph{The set of constraints }$\phi ^{r}\approx 0$ \emph{is regular if and
only if the Jacobian }$J_{i}^{r}\equiv \left. \frac{\partial \phi ^{r}}{%
\partial z^{i}}\right| _{\Sigma }$ \emph{has maximal rank: }$\Re (J)=R$%
.\medskip

A simple classical mechanical example of functionally\emph{\ dependent}
constraints occurs in a $2$-dimensional phase space with coordinates $\left(
q,p\right) $ and constraints $\phi ^{1}\equiv q\approx 0$ and $\phi
^{2}\equiv pq\approx 0$. In this case, $J=\left[
\begin{array}{ll}
1 & p \\
0 & q
\end{array}
\right] _{\Sigma }$ and $\Re \left[ \frac{\partial \left( \phi ^{1},\phi
^{2}\right) }{\partial \left( q,p\right) }\right] _{q=0}=1$. A system of
just one constraint can also fail the test of regularity. Consider the
constraint $\phi =q^{2}\approx 0$ in a $2$-dimensional phase space. In this
case, $J=\left[
\begin{array}{l}
2q \\
0
\end{array}
\right] _{q^{2}=0}=0$ and $\Re \left( J\right) =0$. The same problem occurs
with the constraint $q^{7}\approx 0$, which has a zero of seventh order at
the constraint surface, or with any other\textbf{\ }constraint which is not
linear in $z^{i}-\bar{z}^{i}$.


\section{Classification of Irregular Constraints}

Irregular constraints can be classified according to their approximate
behavior near the surface $\Sigma $.

\textbf{A. Linear constraints}. The Jacobian has constant, non-maximal rank
throughout $\Sigma $, and
\begin{equation}
\phi ^{r}\equiv J_{i}^{r}\left( z^{i}-\bar{z}^{i}\right) \approx 0,\qquad
\Re (J)=R^{\prime }<R.
\end{equation}
These are regular systems in disguise. Regularity fails simply because $%
R-R^{\prime }$ constraints are redundant and should be discarded. The
regular system gives the correct description.

\textbf{B. Multilinear constraints}. In the vicinity of $\Sigma $, the
constraints are of the form
\begin{equation}
\phi \equiv \frac{1}{m!}\,S_{i_{1}\ldots i_{m}}\left( z^{i_{1}}-\bar{z}%
^{i_{1}}\right) \cdots \left( z^{i_{m}}-\bar{z}^{i_{m}}\right) \approx
0\qquad \left( m\leq 2n\right) ,
\end{equation}
where the coefficients $S_{i_{1}\ldots i_{m}}$ vanish if any two indices are
equal. Thus, $\phi $ has \emph{simple zeros} on surfaces of dimension $2n-1$%
, and zeros of higher order occur at the intersections of these surfaces.
The RCs fail at the points of intersection, where $\phi $ has multiple
zeros. At those intersections, $\phi $ can be replaced by the equivalent%
\footnote{%
Two sets of constraints are said to be \emph{equivalent} if they define the
same constraint surface $\Sigma $.} set of constraints,
\begin{equation}
\varphi ^{1}\equiv z^{i_{1}}-\bar{z}^{i_{1}}\approx 0,\qquad \cdots \qquad ,%
\varphi ^{m}\equiv z^{i_{m}}-\bar{z}^{i_{m}}\approx 0.
\end{equation}
At the intersections, the set $\left\{ \varphi ^{1},...,\varphi ^{m}\right\}
$ is regular and therefore this provides a recipe for substituting the
irregular multilinear constraint $\phi $ by a regular set of linear
constraints. For example, the constraint $\phi \equiv qp\approx 0$ is
irregular at $(0,0)$ because it admits a linear approximation everywhere
except at this point. Replacement $\phi $ by the linear constraints $\left\{
q\approx 0,p\approx 0\right\} $ at $(0,0)$ regularizes the system.

\textbf{C. Higher order constraints. }In the vicinity of $\Sigma $, the
constraints do not possess a linear approximation:
\begin{equation}
\phi \equiv C\left( z^{s}-\bar{z}^{s}\right) ^{k}\approx 0\qquad \left(
k>1\right) .  \label{high-order}
\end{equation}
The Jacobian vanishes everywhere on the constraint surface. Naively, it
would seem possible to choose $z^{s}-\bar{z}^{s}\approx 0$ as an equivalent
regular constraint, but it turns out that this could change the dynamics of
original theory, as we show below.

These three classes are generic and, in general, there can be combinations
of these three types occurring simultaneously near a constraint surface. By
selecting a sufficiently small neighborhood of a point on $\Sigma $, one can
always expect to be in one and only one of the three situations just
described.

In correspondence with the three generic cases in which regularity can fail,
the nature of the constraint surface $\Sigma $ falls into one of three
categories:\medskip

\textbf{A. }\emph{The RCs are satisfied on the whole constraint surface. }%
These are regular systems, either desguised or not.

\textbf{B. }\emph{The RCs fail on a submanifold of }$\Sigma $: $\Re \left(
J\right) =R$ except on a submanifold $\Sigma _{0}\subset \Sigma $ where $\Re
\left( J\right) =R^{\prime }<R$\ .

\textbf{C. }\emph{The RCs fail everywhere on }$\Sigma $\emph{:} $J$ has
constant lower than $R$ rank on $\Sigma $.\medskip

The first case is treated in the standard texts and will not be further
discussed here.

In the second case, the constraint surface can be decomposed into two
non-empty submanifolds, $\Sigma _{0}$ and $\Sigma _{R}$ such that $\Sigma
=\Sigma _{0}\cup \Sigma _{R}$ and $\Sigma _{0}\cap \Sigma _{R}$ is empty.
Then, the rank of the Jacobian jumps from $\Re \left( J\right) =R$ on $%
\Sigma _{R}$, to $\Re \left( J\right) =R^{\prime }$ on $\Sigma _{0}$. As
mentioned above, in this case it is possible to replace $\phi $ at $\Sigma
_{0}$ by a set of regular constraints $\left\{ \varphi ^{1}\approx 0,\cdots ,%
\varphi ^{m}\approx 0\right\} $ which regularize the system at $%
\Sigma _{0}$.

The important question is how to proceed in the third case. If the RCs fail
everywhere on $\Sigma $, the previous approach is not applicable, because
there is no guarantee that the resulting Hamiltonian dynamics will be
equivalent to that of the original Lagrangian system.

This can be seen in the example of Lagrangian in a three-dimensional
configuration space,
\begin{equation}
L\left( x,y,z\right) =\dot{x}\dot{z}+yz^{2}.  \label{three particles}
\end{equation}
The general solution of Euler-Lagrange equations\ describes a system with
\emph{one} degree of freedom -- a free particle,\ whose time evolution $\bar{%
x}=p_{0}\,t+x_{0}$ is determined by two initial conditions $p_{0}$\ and $%
x_{0}$. The remaining fields are $\bar{y}(t)$, a Lagrange multiplier with
indeterminate evolution, and $\bar{z}(t)$, with trivial evolution, $\bar{z}%
(t)=0$.

The Hamiltonian approach gives just two (first class) constraints, $\
p_{y}\approx 0$,$\;z^{2}\approx 0$ ($R=2$), and \emph{one} degree of
freedom, as expected from the Lagrangian approach. However, this is only
superficially correct, because the Jacobian has rank 1, and not 2. This is
because the constraint function $\phi =z^{2}$ has no linear approximation at
$z=0$.

On the other hand, if we naively take $z\approx 0$, which is regular and
equivalent to $z^{2}\approx 0$, then the Hamiltonian analysis generates
\emph{three} first class constraints $p_{y}\approx 0,\;z\approx
0,\;p_{x}\approx 0$, which leave \emph{zero} physical degrees of freedom.
This result is not consistent with the Lagrangian description where there is
one degree of freedom.

As seen in the previous example, even if two constraints are equivalent in
defining the same constraint surface, they may yield different dynamics and
should be treated more carefully.

Suppose $\phi \approx 0$ is a constraint of the form (\ref{high-order}),
which is equivalent to the \emph{regular} constraint
\begin{equation}
\chi \equiv \phi ^{1/k}\approx 0.
\end{equation}
The question is whether $\chi \approx 0$ gives also the correct dynamics.
The answer to this question depends on whether $\chi $ is a first or second
class constraint. Namely, it makes a difference whether the linear
constraint $\chi $ can generate a transformation in phase space that leaves
the Hamiltonian action unchanged or not . As shown in \cite{Miskovic-Zanelli}%
, if the linearized constraint $\chi $ is second class, then it is not only
geometrically equivalent to $\phi $ in the sense that it defines the same $%
\Sigma $, but the substitution also yields the same dynamical description as
the Lagrangian approach. On the other hand, if $\chi $ is first class, then
the subtitution generates a system whose dynamics is different from the one
obtained from the Euler-Lagrange equations.


\section{Conclusions}

The recipe for treating the non-regular constraints is:

\begin{itemize}
\item  Every linear or multi-linear set of constraints can be exchanged by
an equivalent regular set. It allows to carry out Dirac's procedure in the
standard way.

\item  A higher order constraint $\phi =C\left( z^{s}-\bar{z}^{s}\right)
^{k}\approx 0$, can be exchanged by the equivalent linear constraint $\chi
=\phi ^{1/k}\approx 0$. If $\chi $ is a second class constraint, the
dynamics of the new system is equivalent to the Lagrangian one. If $\chi $
is first class, the substitution yields a system which is not dynamically
equivalent to the Lagrangian one. In this latter case, one should view the
original Lagrangian as an incomplete, if not a totally inconsistent
description for a dynamical system.
\end{itemize}


\section*{Acknowledgments}

We are deeply grateful to Ricardo Troncoso for many enlightening insights and discussions. This
work is partially funded by grants FONDECYT  1020629, 7020629, 1010450, 2010017, and MECESUP USA
9930 scholarship. The generous support of Empresas CMPC to the CECS is also acknowledged. CECS is
a Millennium Science Institute and it is funded in part by grants from Fundaci\'{o}n Andes and the
Tinker Foundation.



\begin{thebibliography}{9}

\bibitem{Dirac}  P. A. M. Dirac, \emph{Lectures on Quantum Mechanics}
(Yeshiva University, New York, 1964).

\bibitem{Henneaux-Teitelboim}  M. Henneaux and C. Teitelboim, \emph{%
Quantization of Gauge Systems} (Princeton University Press, Princeton, 1992).

\bibitem{Blagojevic}  M. Blagojevi\'{c}, \emph{Gravity and Gauge Symmetries}
(Institute of Physics Publishing, London, 2001).

\bibitem{Chitaia-Gogilidze-Surovtsev}  N. P. Chitaia, S. A. Gogilidze and
Yu. S. Surovtsev, \emph{Phys. Rev.} \textbf{D 56} (1997) 1135; \emph{Phys.
Rev.} \textbf{D 56} (1997) 1142.

\bibitem{Miskovic-Zanelli}  O. Mi\v {s}kovi\'{c} and J. Zanelli, \emph{Dynamical Structure of Irregular Constrained Systems} (in preparation).
\end{thebibliography}
\end{document}